\documentclass[12pt]{iopart}
\usepackage{graphicx}
\usepackage{bm}

\def\re{ \,\mathrm{Re}\,}

\def\bq{ \bm{q} }

\begin{document}

\title{Effect of disorder on the far-infrared conductivity and on the microwave conductivity 
of two-band superconductors}

\author{Bo\v zidar Mitrovi\' c}

\address{Department of Physics, Brock University,
St.Catharines, Ontario, Canada L2S 3A1}
\ead{mitrovic@brocku.ca}

\begin{abstract}
We consider the far-infrared and the microwave conductivities of a two-band superconductor
with non-magnetic impurities. The strong coupling expressions for the frequency 
and temperature dependent conductivity of a two-band superconductor are developed
assuming isotropic bands and interactions. 
Our numerical results obtained using realistic interaction parameters for MgB$_{2}$ are 
compared with experiments on this compound. We find that the available experimental 
results for the far-infrared conductivity of MgB$_{2}$ are consistent with multi-band 
superconductivity in the presence of a sufficiently strong interband impurity scattering.  
On the other hand, our numerical results for the microwave conductivity in the superconducting 
state indicate that the experimental results obtained on samples with the highest 
transition temperature $T_{c}$ are consistent with a low interband impurity scattering rate 
but depend sensitively on the ratio of the total scattering rates in the two bands. For 
the $\pi$-band scattering rate $\gamma_{\pi}$ not greater than the $\sigma$-band scattering rate  
$\gamma_{\sigma}$ there is a single, broad, low-temperature (at about 0.5$T_{c}$) coherence
peak in the microwave conductivity. For $\gamma_{\pi}/\gamma_{\sigma}$=4--7 a high-temperature 
(at about 0.9$T_{c}$) coherence peak is dominant, but there is also a low-temperature peak/shoulder 
resulting from the contribution of the $\pi$-band carriers to the microwave conductivity. For 
$\gamma_{\pi}/\gamma_{\sigma}\gg$1 only the high-temperature coherence peak should be observable. 
\end{abstract}


\submitto{\JPCM}

\maketitle

\section{Introduction}
The far-infrared spectroscopy of superconducting MgB$_{2}$ seems to indicate the presence of a   
single gap $\Delta$ with the ratio 2$\Delta/k_{B}T_{c}$ significantly below the BCS value of 3.53 
(see a recent review by Kuzmenko \cite{Kuzmenko} and the references therein). In contrast to this finding,   
the angle-resolved photoemission spectroscopy \cite{Souma1,Souma2}, the tunneling spectroscopy 
\cite{Giubileo,Szabo,Schmidt,Takasaki1,Iavarone,Gonnelli,Eskildsen,Martinez,Takasaki2}, 
the Raman spectroscopy \cite{Chen,Quilty1,Quilty2,Blumberg} and the heat capacity measurements
\cite{Wang1,Bouquet1,Bouquet2,Wang2} on magnesium diboride have all established two distinct   
superconducting gaps $\Delta_{\sigma}$ and $\Delta_{\pi}$ in $\sigma$- and $\pi$-bands.  
Since the far-infrared measurements \cite{Kaindl,Pimenov} were performed on films with reduced   
superconducting transition temperature $T_{c}$, we investigate the possibility that the interband 
non-magnetic impurity scattering, which is known to reduce the transition temperature of multi-band  
superconductors \cite{Golubov1,Mitrovic1,Nicol,Dolgov}, is responsible for the observation of a single 
gap. Namely, Schopohl and Scharnberg \cite{Schopohl} found that the interband impurity scattering 
produces a common low temperature gap $\Delta$ in both bands of a two-band superconductor such that  
the quasiparticle densities of states in each band vanish at energies below $\Delta$ in the
low temperature limit, figure 1. In the limit of small interband impurity scattering rate, 
the common gap is just above the lower of the two gaps in the clean system (figure 1a). 
As the interband impurity scattering rate increases the common gap grows in size (figure 1b),  
but remains lower than the larger of the two gaps in the clean system, and for a sufficiently   
large interband impurity scattering rate the order parameters (the gap functions) in both  
bands become the same (the Anderson theorem \cite{Anderson}). The growth of the common gap with  
the interband impurity scattering rate is accompanied by progressive smearing of the peaks in 
the quasiparticle densities of states in the two bands, figure 1. Since the frequency ($\nu$)  
dependent conductivity $\sigma(\nu)$ is a functional of both the normal and the anomalous 
quasiparticle densities of states, one would expect the frequency dependence of $\sigma(\nu)$  
to be modified compared to what is predicted by a straightforward application of the Mattis-Bardeen  
theory \cite{Mattis} which uses the BCS form for the quasiparticle densities of states. In particular,  
terahertz time-domain measurements of Kaindl {\em et al.} on MgB$_{2}$ films \cite{Kaindl} 
found that the real part of $\sigma(\nu)$ increased more slowly for $\nu$ just above twice the gap 
than what was predicted by the Mattis-Bardeen theory. This finding is significant since the 
BCS-type Mattis-Bardeen theory describes quite well the low frequency dependence of $\re \sigma(\nu)$  
even for a strong coupling superconductor Pb (see figure 3 in \cite{Palmer}). We find 
that the observed $\sigma(\nu)$ of MgB$_{2}$ films \cite{Kaindl} is consistent with multi-band 
superconductivity in this compound in the presence of a sufficiently strong interband impurity 
scattering.

The microwave conductivity of MgB$_{2}$ was measured in \cite{Jin,Lee,Gennaro} with different results.
In \cite{Jin}, the normalized real part of the conductivity at 17.9 GHz ($\nu$=0.074 meV) measured on
the $c$-axis oriented MgB$_{2}$ films had a broad maximum at a temperature of about 0.6$T_{c}$,
instead of the usual coherence peak seen in the dirty limit just below the $T_{c}$ \cite{Klein}.
The data of Lee {\em et al.} \cite{Lee} obtained at 8.5 GHz ($\nu$=0.035 meV) 
on polycrystalline MgB$_{2}$ films with the $T_{c}$s of 39.3 K and 36.3 K showed rather 
sharp coherence peaks at about 0.9$T_{c}$. In addition, the results obtained on the film with the 
higher $T_{c}$ showed a second, lower, peak at about 0.53$T_{c}$ which appears to be sharper  
than the broad oval feature seen in \cite{Jin} (see figures 3 and 4 in \cite{Lee}). Upon 
ion-milling the second peak became more of a shoulder at about 0.6$T_{c}$, the peak at 0.9$T_{c}$   
became broadened, but increased in size, and the $T_{c}$ dropped from 39.4 K to 36.3 K. The results    
for the normalized microwave conductivity obtained in \cite{Gennaro} at 19 GHz ($\nu$=0.079 meV) 
on MgB$_{2}$ and Mg$_{0.95}$Al$_{0.05}$B$_{2}$ pellets resembled more those found in \cite{Lee}. The 
Al-doped sample displayed more pronounced shoulder at about 0.4$T_{c}$-0.5$T_{c}$ and a more
pronounced coherence peak just below the $T_{c}$ than the magnesium diboride sample. Here we also 
investigate theoretically the effect of impurity scattering on the temperature ($T$) dependence 
of the microwave conductivity $\re \sigma(\nu,T)$ of a two-band superconductor. 
We find that the shape of $\re \sigma(\nu,T)$ depends strongly on both the intraband and the 
interband impurity scattering rates.

The rest of the paper is organized as follows. In section 2 we write down the equations for the
optical conductivity of each band which include explicitly and implicitly the intraband and the 
interband impurity scattering rates. Section 3 contains our results for $\sigma(\nu,T)$ 
obtained using realistic interaction parameters for MgB$_{2}$ and a comparison of experimental 
results \cite{Kaindl} with our theoretical predictions. The same section also contains our results  
for the temperature dependence of the microwave conductivity and their comparison
with experiments \cite{Jin,Lee,Gennaro}. Section 4 contains a summary. In the Appendix we provide 
a connection between 
the main results of \cite{Sung} which used the BCS treatment of the same problem at zero 
temperature and the present work. 

\section{Theory}
The optical conductivity of a two-band superconductor with non-magnetic impurities was 
first considered by Sung and Wong \cite{Sung}. They used the BCS treatment of the pairing 
interactions in two isotropic bands and included the $s$-wave impurity scattering 
in the self-consistent second Born approximation. 
The general expression for the zero temperature optical conductivity of a two-band superconductor
with impurities was developed using the standard linear response theory 
in the conserving approximation \cite{Schrieffer}. Since the electron self-energies were 
obtained in the self-consistent second Born approximation, only the ladder impurity 
diagrams had to be considered in evaluating the current-current correlator 
$-\langle T_{\tau}(j_{\alpha}(\bq=0,-i\tau)j_{\beta}(\bq=0,0))$ \cite{Schrieffer}.  
However, the graphs where the interband scattering impurity lines cross the current vertex vanish 
because of the assumed isotropy of the bands and impurity scattering matrix elements \cite{Rainer}. 
As a result, the paramagnetic part of the electromagnetic  
response kernel is simply the sum of the contributions from individual bands. 
In terms of concrete quantitative predictions, they only pointed out that  
the absorption threshold should increase with increasing interband impurity scattering 
rate/impurity concentration, figure 1.

Here we generalize the results in \cite{Sung} to include the strong-coupling effects because it is known
\cite{Mitrovic1},\cite{Nicol},\cite{Dolgov} that the weak-coupling BCS treatment of the effect of 
interband impurity scattering \cite{Golubov1} badly overestimates the rate of suppression of $T_{c}$ 
in MgB$_{2}$. Also, the BCS treatment requires an unphysical value of the Debye cutoff of 7.5 meV 
\cite{Lee}, while it is known that the superconductivity of MgB$_{2}$ is mainly driven by electron 
coupling to the optical B-B bond stretching modes at about 74 meV. 

As in \cite{Sung}, we will make the assumption of isotropic bands and interactions (electron-phonon, 
screened Coulomb and electron-impurity). The assumed isotropy of the interaction parameters implies 
that there are no vertex corrections to current vertices in the current-current correlator 
$-\langle T_{\tau}(j_{\alpha}(\bq=0,-i\tau)j_{\beta}(\bq=0,0))$ 
\cite{Rainer} and, as in \cite{Sung}, the conductivity is simply the sum of conductivities of 
separate bands. The calculation of the $\bq=0$ conductivity $\sigma_{n}(\nu)$ of a band $n$ 
in the superconducting state 
is performed in the standard way (see, for example, a pedagogical account in \cite{Marsiglio} and 
the references therein) and one finds the usual result \cite{Klein,Marsiglio}

\begin{eqnarray}
\label{conductivity}   
\fl \sigma_{n}(\nu) =  \frac{\omega_{pn}^{2}}{8\pi\nu}\left\{ 
\int_{0}^{+\infty}d\omega\tanh\frac{\omega}{2T}\;
\frac{1-N_{n}(\omega)N_{n}(\omega+\nu)-M_{n}(\omega)M_{n}(\omega+\nu)}{
-iE_{n}(\omega)-iE_{n}(\omega+\nu)} \right. 
                                                                     \nonumber \\
\fl \phantom{\sigma_{n}(\nu)}+\int_{0}^{+\infty}d\omega\tanh\frac{\omega+\nu}{2T}\;
\frac{1-N_{n}^{*}(\omega)N_{n}^{*}(\omega+\nu)-M_{n}^{*}(\omega)M_{n}^{*}(\omega+\nu)}{
-iE_{n}^{*}(\omega)-iE_{n}^{*}(\omega+\nu)}
                                                                      \nonumber \\
\fl \phantom{\sigma_{n}(\nu)}
+\int_{0}^{+\infty}d\omega\left(\tanh\frac{\omega+\nu}{2T}-\tanh\frac{\omega}{2T}\right)\;
\frac{1+N_{n}^{*}(\omega)N_{n}(\omega+\nu)+M_{n}^{*}(\omega)M_{n}(\omega+\nu)}{
iE_{n}^{*}(\omega)-iE_{n}(\omega+\nu)}
                                                                     \nonumber \\
\fl \phantom{\sigma_{n}(\nu)}
+\int_{-\nu}^{0}d\omega\tanh\frac{\omega+\nu}{2T}\; 
\left[\frac{1-N_{n}^{*}(\omega)N_{n}^{*}(\omega+\nu)-M_{n}^{*}(\omega)M_{n}^{*}(\omega+\nu)}{
-iE_{n}^{*}(\omega)-iE_{n}^{*}(\omega+\nu)} \right.
                                                                      \nonumber \\
\fl \phantom{\sigma_{n}(\nu)}
  \left. \left. +\frac{1+N_{n}^{*}(\omega)N_{n}(\omega+\nu)+M_{n}^{*}(\omega)M_{n}(\omega+\nu)}{
iE_{n}^{*}(\omega)-iE_{n}(\omega+\nu)}
\right]\right\}\>,
\end{eqnarray}
where $n,m=\sigma,\pi$. Here, $\omega_{pn}^{2}$ is the square of the plasma 
frequency in the band n, and the real parts of 
\begin{equation}
\label{DOS}
N_{n}(\omega)=\frac{\omega}{\sqrt{\omega^{2}-\Delta_{n}^{2}(\omega)}}
\end{equation}
and 
\begin{equation}
\label{ADOS}
M_{n}(\omega)=\frac{\Delta_{n}(\omega)}{\sqrt{\omega^{2}-\Delta_{n}^{2}(\omega)}}
\end{equation}  
are normalized normal and anomalous quasiparticle densities of states, respectively, in the band $n$. 
The quasiparticle energy in the band $n$, 
$E_{n}(\omega)$, appearing in the denominators in (\ref{conductivity}) is defined by 
\begin{equation}
\label{ENERGY}
E_{n}(\omega)=Z_{n}(\omega)\sqrt{\omega^{2}-\Delta_{n}^{2}(\omega)}\>,
\end{equation}
where $Z_{n}(\omega)$ is the total renormalization function for the band $n$ which includes 
the intraband and the interband electron-phonon interaction and impurity scattering, 
and $\Delta_{n}(\omega)$ is the total gap function (renormalized pairing self-energy \cite{Schrieffer})
in the band $n$ 
which depends on the intraband and the interband electron-phonon interaction and screened 
Coulomb interaction and only on the interband impurity scattering (see below). In (\ref{DOS}-
\ref{ENERGY}) and throughout this paper we take the branch of the square root with positive 
imaginary part and the energy $\omega$ is assumed to have infinitesimal positive imaginary part. 

The gap functions $\Delta_{n}(\omega)$ and the renormalization functions $Z_{n}(\omega)$ are 
obtained by solving the Eliashberg equations at finite temperature on the real axis \cite{Mitrovic2}
\begin{equation}
\label{Eli1}
\phi_{n}(\omega)=\phi_{n}^0(\omega)+i\sum_{m}\frac{1}{2\tau_{nm}}
  \frac{\Delta_{m}(\omega)}{\sqrt{\omega^{2}-\Delta_{m}^{2}(\omega)}}\>,
\end{equation}
\begin{eqnarray}
\label{Eli2}
 \phi_{n}^0(\omega)=\sum_{m}\int\limits_0^{\omega_c}d\omega'\re[M_{m}(\omega)]\left[ 
f(-\omega')K_{nm}^{+}(\omega,\omega')-f(\omega')K_{nm}^{+}(\omega,-\omega')\right. \nonumber \\
 \phantom{\phi_{n}^0(\omega)=}\left. -\mu_{nm}^{*}(\omega_c)\tanh\frac{\omega'}{2T} 
+{\bar K}_{nm}^{+}(\omega,\omega')-{\bar K}_{nm}^{+}(\omega,-\omega')\right]\>,
\end{eqnarray}
\begin{equation}
\label{Eli3}
Z_{n}(\omega)=Z_{n}^0(\omega)+i\sum_{m}\frac{1}{2\tau_{nm}}
\frac{1}{\sqrt{{\omega}^2-\Delta_m^2(\omega)}}\>,
\end{equation} 
\begin{eqnarray}
\label{Eli4}
Z_{n}^0(\omega)=1-\frac{1}{\omega} \sum_m\int\limits_0^{+\infty}d\omega'\re[N_{m}(\omega)]\left[
f(-\omega')K_{nm}^{-}(\omega,\omega')  \right. \nonumber \\
 \phantom{Z_{n}^0(\omega)=}\left. -f(\omega')K_{nm}^{-}(\omega,-\omega')+
{\bar K}_{nm}^{-}(\omega,\omega')+{\bar K}_{nm}^{-}(\omega,-\omega') 
\right]\>.
\end{eqnarray}
Equations (\ref{Eli1}-\ref{Eli4}) imply that the gap functions $\Delta_{n}(\omega)=
\phi_{n}(\omega)/Z_{n}(\omega)$ are independent of the intraband impurity scattering rate 
$1/\tau_{nn}$, but are affected by the interband impurity scattering rate $1/\tau_{nm}$, $n\neq m$.  
The intraband and the interband electron-phonon coupling functions $\alpha^{2}F_{nm}(\Omega)$
enter via the zero temperature kernels $K_{nm}^{\pm}(\omega,\omega')$ and the 
thermal phonon kernels ${\bar K}_{nm}^{\pm}(\omega,\omega')$ defined by
\begin{equation}
\label{kernel}
\fl K_{nm}^{\pm}(\omega,\omega')=\int\limits_0^{+\infty}d\Omega\alpha^{2}F_{nm}(\Omega)
\left[\frac{1}{\omega'+\omega+\Omega+i0^{+}}\pm  \frac{1}{\omega'-\omega+\Omega-i0^{+}}\right]\>,
\end{equation}
\begin{equation}
\label{TPkernel}
\fl {\bar K}_{nm}^{\pm}(\omega,\omega')=\int\limits_0^{+\infty}d\Omega
\frac{\alpha^{2}F_{nm}(\Omega)}{e^{\Omega/T}-1}
\left[\frac{1}{\omega'+\omega+\Omega+i0^{+}}
\pm   \frac{1}{\omega'-\omega+\Omega-i0^{+}}\right]\>.
\end{equation}
The screened Coulomb interaction for the cutoff $\omega_{c}$ in the Eliashberg equations 
is described by the intraband and the interband Coulomb repulsion parameters 
$\mu_{nm}^{*}(\omega_c)$.
The impurity scattering rates are defined by $\gamma_{nm}\equiv1/\tau_{nm}=
2\pi n_{imp}N_{Fm}|V_{nm}|^2$, where $n_{imp}$ is the concentration of impurities, 
$N_{Fm}$ is the normal state electronic density of states per spin at the Fermi level in band $m$  
and $V_{nm}$ is the Fermi surface averaged
matrix element of the change in the lattice
potential caused by an impurity between the states
in the bands $n$ and $m$.

In the single band case it is customary to make the dependence of the conductivity on the 
impurity scattering rate more explicit as it enters only the total renormalization function \cite{Klein}. 
The same can be done in multi-band case, while keeping in mind that the interband impurity 
scattering also enters implicitly through the gap function $\Delta_{n}(\omega)$, equations 
(\ref{conductivity}-\ref{ENERGY}). To this end $Z_{n}(\omega)$ in (\ref{ENERGY}) 
is replaced by (\ref{Eli3}) and one has 
\begin{eqnarray}
\label{ENERGY1}
E_{n}(\omega)=Z_{n}^{0}(\omega)\sqrt{\omega^{2}-\Delta_{n}^{2}(\omega)}+\frac{i}{2\tau_{nn}}+
\frac{i}{2\tau_{nm}}\frac{\sqrt{\omega^{2}-\Delta_{n}^{2}(\omega)}}{\sqrt{\omega^{2}-\Delta_{m}^{2}(\omega)}}
                                                                \\
\label{ENERGY2}
\phantom{E_{n}(\omega)}=Z_{n}^{0}(\omega)\sqrt{\omega^{2}-\Delta_{n}^{2}(\omega)}+\frac{i}{2\tau_{nn}}
+\frac{i}{2\tau_{nm}}\frac{N_{m}(\omega)}{N_{n}(\omega)}
\end{eqnarray}
with $m\neq n$, and in the second line we have utilized the definition (\ref{DOS}). 
We point out that in the limit of zero interband scattering $1/\tau_{nm}$=0, $m\neq n$,
(\ref{ENERGY2}) is given by
$E_{n}(\omega)=Z_{n}^{0}(\omega)\sqrt{\omega^{2}-\Delta_{n}^{2}(\omega)}+i/2\tau_{nn}$ and
(\ref{conductivity}) takes the usual single band form given in \cite{Klein} (see equation (3) in
\cite{Klein}). 

In the normal state  $N_{n}(\omega)=$1, $M_{n}(\omega)=$0, $E_{n}(\omega)=Z_{n}^{0}(\omega)\omega$ and 
(\ref{conductivity}) reduces to 
\begin{eqnarray}
\label{nconductivity}
\fl \sigma_{n}^{N}(\nu) =  \frac{\omega_{pn}^{2}}{8\pi\nu}\left\{
\int_{0}^{+\infty}d\omega\left(\tanh\frac{\omega+\nu}{2T}-\tanh\frac{\omega}{2T}\right)\times \right.
                                                               \nonumber \\
\fl \phantom{\sigma_{n}^{N}(\nu)=}\frac{2}{i{Z_{n}^{0}}^{*}(\omega)\omega-iZ_{n}^{0}(\omega+\nu)
(\omega+\nu)+1/\tau_{nn}+1/\tau_{nm}}   \nonumber \\
\fl \phantom{\sigma_{n}^{N}(\nu)=}  \left. +\int_{-\nu}^{0}d\omega\tanh\frac{\omega+\nu}{2T}
\frac{2}{i{Z_{n}^{0}}^{*}(\omega)\omega-iZ_{n}^{0}(\omega+\nu)(\omega+\nu)+
1/\tau_{nn}+1/\tau_{nm}} \right\}\>,
\end{eqnarray}
with $m\neq n$, which is the standard result for the total impurity scattering rate $\gamma_n=
1/\tau_{n}=1/\tau_{nn}+1/\tau_{nm}$, $m\neq n$, in the band $n$.
In (\ref{nconductivity}) $Z_{n}^{0}(\omega)$ is obtained from (\ref{Eli4}) with $N_{m}(\omega)$=1.

In the Appendix we provide a translation of the notation used in \cite{Sung} 
into the notation used in this work in order to facilitate  comparison between our results and 
the BCS zero-temperature results of \cite{Sung}.  

\section{Numerical Results}
In all of our numerical work we use four electron-phonon coupling functions 
$\alpha^{2}F_{nm}(\Omega)$, $n,m=\sigma,\pi$, for MgB$_{2}$ calculated by Golubov {\em et al.} 
\cite{Golubov2}. 
The corresponding electron-phonon coupling parameters $\lambda_{nm}=\re K_{nm}^{+}(0,0)$, equation 
(\ref{kernel}), are $\lambda_{\sigma\sigma}$ = 1.017, $\lambda_{\sigma\pi}$ = 0.212, 
 $\lambda_{\pi\pi}$ = 0.446 and $\lambda_{\pi\sigma}$ = 0.155. The  Coulomb repulsion parameters 
$\mu_{nm}^*(\omega_c)$ were determined in \cite{Mitrovic1} 
based on the screened Coulomb interactions in 
MgB$_{2}$ calculated by Choi {\em et al.} \cite{Choi} by fitting to the 
experimental $T_c$ of the clean system. The solutions $\Delta_{\sigma}(\omega)$ and 
$\Delta_{\pi}(\omega)$ of (\ref{Eli1}-\ref{Eli4}) depend only on the band off-diagonal impurity 
scattering rates and since $\gamma_{\sigma\pi}/\gamma_{\pi\sigma}=N_{F\pi}/N_{F\sigma}=
\lambda_{\sigma\pi}/\lambda_{\pi\sigma}=$1.37, there is only one independent interband 
scattering parameter and we chose $\gamma_{\pi\sigma}$ as the independent parameter. 

The contributions to the conductivity 
of the carriers in the two bands depend also on the intraband impurity scattering rates and 
on the plasma frequencies of the two bands and in the following we choose the values of 
$\gamma_{\sigma\sigma}=1/\tau_{\sigma\sigma}$, $\gamma_{\pi\pi}=1/\tau_{\pi\pi}$,  
$\omega_{p\sigma}$ and $\omega_{p\pi}$ according to 
a particular set of experiments, guided by the band structure calculations. However, 
it should be noted that there is no unique way of determining even just two parameters 
$\gamma_{\sigma\sigma}$ and $\gamma_{\pi\pi}$ from the measured conductivity just above the 
$T_{c}$, assuming that the plasma frequencies are as given by the band structure 
calculations and that the interband scattering rate $\gamma_{\pi\sigma}$ can be deduced 
from the $T_{c}$ of the film and the calculated $T_{c}$ vs.~$\gamma_{\pi\sigma}$ curve 
\cite{Mitrovic1,Nicol,Dolgov}.
This uncertainty necessarily makes any detailed comparison with experiments  
difficult. Thus, we will focus on qualitative changes in the conductivities of the two 
bands brought about by the two-band superconductivity in MgB$_{2}$ in the presence of 
impurity scattering.

\subsection{Frequency and temperature dependence of the far-infrared conductivity}

In the experiments of Kaindl {\em et al.} on MgB$_{2}$ \cite{Kaindl} the $T_{c}$ of the 100 nm film 
used for the measurement of $\sigma(\nu)$ was 30.5 K. If one assumes that this reduction in 
the transition temperature from the maximum value of $T_{c0}$=39.4 K  results solely from the interband 
impurity scattering (i.~e.~if one ignores any possible changes to the Fermi level densities of states 
$N_{F\sigma}$ and $N_{F\pi}$ and/or the electron-phonon coupling functions) one can deduce the value of the  
interband impurity scattering rate $\gamma_{\pi\sigma}$ from the $T_{c}$ vs.~$\gamma_{\pi\sigma}$  
curve calculated in \cite{Mitrovic1}. In this way we find $\gamma_{\pi\sigma}$=5$T_{c0}$, where $T_{c0}$ is
the transition temperature of the clean system.  
The corresponding normalized quasiparticle densities of states obtained from the solutions of 
(\ref{Eli1}-\ref{Eli4}) are shown in figure 1b. The measured real part of the normal state 
conductivity at 40 K was about 8$\times$10$^{5}\Omega^{-1}$m$^{-1}$ (see the inset in figure 2 in   
\cite{Kaindl}).
This value is mainly determined by the impurity scattering and one can deduce 
$\gamma_{\sigma\sigma}$ and $\gamma_{\pi\pi}$ by choosing appropriate values of $\omega_{p\sigma}$ 
and $\omega_{p\pi}$ and making an assumption about the ratio $\gamma_{\pi}/\gamma_{\sigma}$, where
$\gamma_{\pi}=\gamma_{\pi\pi}+\gamma_{\pi\sigma}$ and 
$\gamma_{\sigma}=\gamma_{\sigma\sigma}+\gamma_{\sigma\pi}$. We chose $\omega_{p\pi}$=5.89 eV and 
$\omega_{p\sigma}$=4.14 eV calculated in \cite{Mazin} and assumed $\gamma_{\pi}/\gamma_{\sigma}$=7 as 
suggested by Kuzmenko \cite{Kuzmenko}, which resulted in $\gamma_{\pi}$=2.6 eV and  
$\gamma_{\sigma}$=0.371 eV. These choices do not affect the calculated $\re \sigma_{n}(\nu)/
\re \sigma_{n}^{N}(\nu)$, $n=\sigma,\pi$ given by (\ref{conductivity}) and (\ref{nconductivity}) but 
only $\re \sigma(\nu)/\re \sigma^{N}(\nu)$, with $\sigma(\nu)=\sigma_{\sigma}(\nu)+\sigma_{\pi}(\nu)$. 
In figure 2 we show our results for the normalized conductivities at several temperatures and the 
results obtained by using the single-gap Mattis-Bardeen theory \cite{Mattis} (dashed lines). With our 
interaction parameters described at the beginning of this section and $\gamma_{\pi\sigma}$=5$T_{c0}$ 
we obtained for the common low temperature gap, figure 1b, $\Delta$=3.875 meV. The transition 
temperature $T_{c}$ was determined from the temperature dependence of the conductivity in the 
low frequency limit (see the next subsection) and we found $T_{c}$=33.2 K giving 2$\Delta/k_{B}T_{c}$=
2.7 which is higher than the value found in \cite{Kaindl} but still well below the  
BCS value of 3.53. The five temperatures for which we calculated the conductivities in figure 2 
were chosen such that they correspond to the same values of $T/T_{c}$ considered in \cite{Kaindl} 
and they are $T$=7 K, 19 K, 26 K, 
29 K and 32.75 K. The results of Mattis-Bardeen theory (dashed lines in figure 2) were obtained by 
taking $\Delta$=3.875 meV as the zero temperature gap and assuming that the temperature dependence 
of the gap is given by the BCS theory as calculated by M\" uhlschlegel \cite{Muhlschlegel}.

The most important feature of our results in figure 2 is that both $\re \sigma_{\sigma}(\nu)/
\re \sigma_{\sigma}^{N}(\nu)$ and $\re \sigma_{\pi}(\nu)/\re \sigma_{\pi}^{N}(\nu)$ increase 
more slowly above twice the gap than what is predicted by the Mattis-Bardeen theory, in particular 
at the lower temperatures. The reduction in the rate of increase in the normalized real part of the 
conductivity compared to the prediction of the Mattis-Bardeen theory is more pronounced for the 
$\sigma$-band than for the $\pi$-band. The reduced rate of increase in $\re \sigma_{n}(\nu)/
\re \sigma_{n}^{N}(\nu)$, $n=\sigma,\pi$ is related to the smearing of both the normal quasiparticle  
density of states $\re N_{n}(\omega)$, figure 1, and the anomalous quasiparticle density of states 
$M_{n}(\omega)$ by the interband impurity scattering. Note that for the interband impurity 
scattering rate $\gamma_{\pi\sigma}$=5$T_{c0}$, which was used to obtain the results in figure 2, 
$\re N_{\sigma}(\omega)$ is broadened more than $\re N_{\pi}(\omega)$, figure 1b, resulting in a  
slower increase 
of $\re \sigma_{\sigma}(\nu)/\re \sigma_{\sigma}^{N}(\nu)$ compared to $\re \sigma_{\pi}(\nu)/
\re \sigma_{\pi}^{N}(\nu)$. Indeed, in figure 3 we show $\re \sigma_{\pi}(\nu)/
\re \sigma_{\pi}^{N}(\nu)$ at a low temperature calculated for $\gamma_{\pi\sigma}$=0.1$T_{c0}$, 
together with the prediction of the Mattis-Bardeen theory. For such a small interband 
impurity scattering rate the smearing in $\re N_{\pi}(\omega)$ compared to the BCS result 
$\re( \omega/\sqrt{\omega^2-\Delta^2})$ is quite small, figure 1a,  
and in this case the Mattis-Bardeen theory provides
an excellent fit. We conclude that the observed \cite{Kaindl} single gap and a slower rise in 
$\re \sigma(\nu)/\re \sigma^{N}(\nu)$ above twice the gap compared to the 
prediction of the Mattis-Bardeen theory 
are consistent with the multi-band superconductivity in MgB$_{2}$ in the presence of a 
sufficiently strong interband impurity scattering. 

\subsection{Temperature dependence of the microwave conductivity}

In figure 4 we show the temperature dependence of the microwave conductivity at 
8.5 GHz ($\nu$=0.035 meV) calculated from equation (\ref{conductivity}) using the values of 
$\omega_{p\sigma}$, $\omega_{p\pi}$ and $\gamma_{\pi}/\gamma_{\sigma}$ suggested by Kuzmenko  
\cite{Kuzmenko}: $\omega_{p\sigma}$=4.14 eV, $\omega_{p\pi}$=4.72 eV and $\gamma_{\pi}/\gamma_{\sigma}$=7. 
The value of $\gamma_{\pi}$ was fitted to the microwave conductivity of 1.26$\times$
10$^{7}/\Omega$m measured in \cite{Lee} at the transition temperature of the film of MgB$_2$ with a higher 
$T_{c}$ ($T_{c}$=39.3 K). We obtained $\gamma_{\pi}$=152.6 meV and for different choices of 
$\gamma_{\pi\sigma}$ in figure 4 the values of $\gamma_{\pi\pi}$ and $\gamma_{\sigma\sigma}$ 
were adjusted to keep $\gamma_{\pi}=\gamma_{\pi\pi}+\gamma_{\pi\sigma}$ 
and $\gamma_{\sigma}=\gamma_{\sigma\sigma}+\gamma_{\sigma\pi}$ fixed (note that $\gamma_{\sigma\pi}/
\gamma_{\pi\sigma}$=1.37 is constant). In this way the same microwave 
conductivity at $T_{c}$ is obtained for different values of the interband scattering rate.   

The results in figure 4 are analogous to what was obtained previously by Mitrovi\' c and Samokhin 
\cite{Mitrovic2} for the nuclear magnetic resonance (NMR) relaxation rate in two-band superconductors. 
This is because both the NMR relaxation rate and the microwave conductivity have the same coherence  
factors in the single band case. For no interband impurity scattering ($\gamma_{\pi\sigma}$=0) the 
microwave conductivity of the $\sigma$-band has the usual coherence peak at about 0.9$T_{c}$, while 
the microwave conductivity of the $\pi$-band displays an unusual broad peak, first noted in \cite{Jin} 
at about 0.4$T_{c}$-0.5$T_{c}$. The difference in temperatures of the two coherence peaks is a direct 
consequence of the difference in the energies at which the low temperature quasiparticle densities of   
states in the two bands have singularities, figure 1a. For finite interband impurity scattering, 
the transition temperature is reduced with increasing $\gamma_{\pi\sigma}$ and the size of the 
coherence peak in the $\sigma$-band contribution to the microwave conductivity is reduced for
$\gamma_{\pi\sigma}$ up to about $T_{c0}$, figure 4a, as a result of the reduction and broadening   
of the peaks in $\re N_{\sigma}(\omega)$ and $\re M_{\sigma}(\omega)$. Since the peaks in  
$\re N_{\pi}(\omega)$ and $\re M_{\pi}(\omega)$ are less smeared for low values of $\gamma_{\pi\sigma}$ 
than those in the $\sigma$-band, figure 1a, the effect of the interband impurity scattering on 
a broad $\pi$-band coherence peak is small for small $\gamma_{\pi\sigma}$. As $\gamma_{\pi\sigma}$  
grows, $\re N_{\pi}(\omega)$ and $\re M_{\pi}(\omega)$ become more broadened and start approaching 
$\re N_{\sigma}(\omega)$ and $\re M_{\sigma}(\omega)$, figure 1b, as the difference in the gap functions 
in the two bands becomes smaller. The consequence of these changes in $N_{\pi}(\omega)$ 
and $M_{\pi}(\omega)$ is that the coherence peak in $\pi$-band contribution to the microwave conductivity 
starts moving closer to the $T_{c}$ and the shape of $\re \sigma_{\pi}(T)$ for fixed $\nu$ in the 
microwave range starts resembling that of $\re \sigma_{\sigma}(T)$. In figure 5 we show the microwave 
conductivities calculated for parameters used in the previous subsection with $\gamma_{\pi\sigma}$=5$T_{c0}$. 
The shapes of $\re \sigma_{\pi}(T)$ and $\re \sigma_{\sigma}(T)$ are qualitatively the same since 
the peaks in the corresponding densities of states occur at similar energies, figure 1b. As pointed out 
in \cite{Mitrovic2}, in the limit of very large $\gamma_{\pi\sigma}$ 
(the Anderson limit \cite{Anderson}) the gap functions in the two bands become identical leading 
to identical normal and anomalous quasiparticle densities of states in both bands. This in turn 
would imply the usual temperature dependence of the microwave conductivity with the coherence peak  
at about 0.9$T_{c}$, barring extremely strong electron-phonon coupling \cite{Marsiglio}.

From figure 4 it is clear that the results of Jin {\em et al.} \cite{Jin} with a broad coherence peak at 
about 0.6$T_{c}$ could be obtained with a small interband scattering rate $\gamma_{\pi\sigma}$, 
which is consistent with a rather high $T_{c}$=39.4 K of their samples \cite{Kang}, 
{\em and} with $\gamma_{\pi}$ less than or comparable to $\gamma_{\sigma}$. 
Indeed, in figure 6 we show a series of our results calculated with 
$\gamma_{\pi\sigma}$=0.1$T_{c0}$ and $\gamma_{\pi}$=0.5$\gamma_{\sigma}$ (figure 6a), 
$\gamma_{\pi}$=$\gamma_{\sigma}$ (figure 6b), $\gamma_{\pi}$=2$\gamma_{\sigma}$ (figure 6c), $\gamma_{\pi}$=
2.67$\gamma_{\sigma}$ (figure 6d), $\gamma_{\pi}$=4$\gamma_{\sigma}$ (figure 6e) and $\gamma_{\pi}$=
6.67$\gamma_{\sigma}$ (figure 6f). The dashed curves in figure 6 give the $\sigma$-band contributions to 
the microwave conductivity, dash-dotted curves give the $\pi$-band contributions to the microwave
conductivity and the solid lines give the total microwave conductivity. We used $\omega_{p\pi}$=5.89 eV, 
$\omega_{p\sigma}$=4.14 eV \cite{Mazin} and the value of $\gamma_{\pi}$ was fitted to the measured 
microwave conductivity at $T_{c}$ of 1.37$\times$10$^{7}/\Omega$m \cite{Jin} assuming 
$\re \sigma_{\sigma}(T_{c})=\omega_{p\pi}^2/(4\pi\gamma_{\pi})+\omega_{p\sigma}^2/(4\pi\gamma_{\sigma})$.  
In this way we obtained $\gamma_{\pi}$=42.5 meV, 50.9 meV, 68.1 meV, 78.6 meV, 100.9 meV and 
145.4 meV  for figures 6a through 6f, respectively. Clearly, only the results in figures 6a and 6b 
are consistent with the experimental observation in \cite{Jin}. The calculated microwave conductivity in 
figures 6e and 6f is consistent with the experimental findings in \cite{Lee,Gennaro}, but rather sharp   
peaks at 0.53$T_{c}$ and 0.9$T_{c}$ observed in \cite{Lee} on a film with the higher $T_{c}$ (39.3 K)  
cannot be reproduced theoretically. We note that the shape of the calculated $\sigma$-band contribution to 
the microwave conductivity is quite similar to that of Nb \cite{Jin,Klein}, which is not surprising 
since the $\sigma$-band electron phonon coupling parameter $\lambda_{\sigma\sigma}$=1.017 is comparable 
to that of Nb.

\section{Summary}

We have developed strong coupling expressions for the frequency and temperature dependent 
conductivity of a two-band superconductor which include the intraband and the interband 
scattering by non-magnetic impurities, assuming isotropic bands and interactions. Our  
numerical calculations, using realistic interaction parameters for MgB$_{2}$, show that the 
experimental observations \cite{Kaindl} of a single gap and a lower rate of increase in the far infrared 
conductivity above the absorption threshold, compared to the prediction of Mattis-Bardeen 
theory \cite{Mattis}, are consistent with multi-band superconductivity in MgB$_{2}$ in the presence of a 
sufficiently strong disorder. The results for the microwave conductivity show that the intraband and 
the interband impurity scattering rates play the key role in determining its temperature dependence.  
The experimental results in \cite{Jin,Lee,Gennaro}, at least on the samples with nearly 
optimum transition temperatures, are consistent with a low interband impurity scattering rates and 
their precise shape seems to depend on the sample quality as reflected by the ratio 
$\gamma_{\pi}/\gamma_{\sigma}$ of the impurity scattering rates in the two bands. For 
$\gamma_{\pi}/\gamma_{\sigma}\leq$1 the theory predicts the temperature dependence of the 
microwave conductivity observed in \cite{Jin}, while the theoretical results obtained for 
$\gamma_{\pi}/\gamma_{\sigma}\approx$4--7 are consistent with the observations in \cite{Lee,Gennaro}. 

\ack
This work has been supported in part by the Natural Sciences and Engineering 
Research Council of Canada. The author is grateful to O.~Jepsen for providing the numerical values 
of $\alpha^{2}F$s for MgB$_{2}$ presented in \cite{Golubov2} and to K.~V.~Samokhin  
for many useful conversations on multi-band superconductors.
\appendix
\section*{Appendix}
\setcounter{section}{1}

The quantities $\tilde{\omega}_{n}$, $\tilde{\Delta}_{n}$, $\Delta_{n}$ and $\Gamma_{nm}$ used in \cite{Sung} 
are expressed in the notation of the present paper as $\tilde{\omega}_{n}=\omega Z_{n}(\omega)$,
$\tilde{\Delta}_{n}=\phi_{n}(\omega)$, $\Delta_{n}=\phi_{n}^{0}$ and $\Gamma_{nm}=1/(2\tau_{nm})$ 
(compare equations (8-11,14,15) in \cite{Sung} with (\ref{Eli1}-\ref{Eli4}) in this 
work). Note that we reserve the notation $\Delta_{n}(\omega)=\phi_{n}(\omega)/Z_{n}(\omega)$ 
for the true physical gap function which is experimentally observable, while in 
\cite{Sung} it denotes the pairing self-energy resulting from pairing interactions (see 
equations (14) and (15) in \cite{Sung}) which is experimentally unobservable for a finite 
iterband impurity scattering rate even in the BCS limit. 
Then, the functions $u_{n}(\omega)$ and
$v_{n}(\omega)$ of \cite{Sung} become
$u_{n}(\omega)\equiv\tilde{\omega}_{n}/\tilde{\Delta}_{n}=\omega/\Delta_{n}(\omega)$
and $v_{n}(\omega)\equiv\phi_{n}^{0}u_{n}(\omega)$ and the factor in the
square bracket under the integral in the equation (38a) of \cite{Sung} can be rewritten as
($u_{-n}\equiv u_{n}(\omega-\nu)$)
\begin{eqnarray}
\fl 1-\frac{v_{n}v_{-n}+{\phi_{n}^{0}}^{2}}{\sqrt{v_{n}^{2}-{\phi_{n}^{0}}^{2}}
\sqrt{v_{-n}^{2}-{\phi_{n}^{0}}^{2}}}
 =1-\frac{u_{n}u_{-n}+1}{\sqrt{u_{n}^{2}-1}\sqrt{u_{-n}^{2}-1}} \nonumber \\
\fl \phantom{1-\frac{v_{n}v_{-n}+{\phi_{n}^{0}}^{2}}{\sqrt{v_{n}^{2}-{\phi_{n}^{0}}^{2}}
\sqrt{v_{-n}^{2}-{\phi_{n}^{0}}^{2}}}}
 = 1-\frac{\omega(\omega-\nu)+\Delta_{n}(\omega)\Delta_{n}(\omega-\nu)}
{\sqrt{\omega^2-\Delta_{n}^{2}(\omega)}{\sqrt{(\omega-\nu)^2-\Delta_{n}^{2}(\omega-\nu)}}}
   \nonumber \\
\label{SW}
\fl \phantom{1-\frac{v_{n}v_{-n}+{\phi_{n}^{0}}^{2}}{\sqrt{v_{n}^{2}-{\phi_{n}^{0}}^{2}}
\sqrt{v_{-n}^{2}-{\phi_{n}^{0}}^{2}}}}
 = 1-[N_{n}(\omega)N_{n}(\omega-\nu)+M_{n}(\omega)M_{n}(\omega-\nu)]\>,
\end{eqnarray}
where in the last step we used the definitions (\ref{DOS}) and (\ref{ADOS}). Moreover, the function
$\gamma_{n}$ given by the equation (38b) in \cite{Sung}, which determines the remaining factor 
$1/(\gamma_{n}+\gamma_{-n})$ in the integrand of (38a) in \cite{Sung}, can be cast in the following form
\begin{eqnarray}
\gamma_{n}=\sqrt{v_{n}^{2}-{\phi_{n}^{0}}^{2}}+i\Gamma_{n}+i\Gamma_{nm}\frac{\phi_{m}^{0}}{\phi_{n}^{0}}
\left(\frac{v_{n}^{2}-{\phi_{n}^{0}}^{2}}{v_{m}^{2}-{\phi_{m}^{0}}^{2}}\right)^{1/2} \nonumber \\
\phantom{\gamma_{n}}=\phi_{n}^{0}\sqrt{u_{n}^{2}-1}+i\Gamma_{n}+i\Gamma_{nm}
\left(\frac{u_{n}^{2}-1}{u_{m}^{2}-1}\right)^{1/2} \nonumber \\
\phantom{\gamma_{n}}=\phi_{n}^{0}\frac{\sqrt{\omega^2-\Delta_{n}^{2}(\omega)}}{\Delta_{n}(\omega)}+i\Gamma_{n}
+i\Gamma_{nm}\frac{\Delta_{m}(\omega)}{\Delta_{n}(\omega)}\frac{\sqrt{\omega^2-\Delta_{n}^{2}(\omega)}}
{\sqrt{\omega^2-\Delta_{m}^{2}(\omega)}} \nonumber \\
\label{SW1}
\phantom{\gamma_{n}}=
\left(\phi_{n}^{0}+i\Gamma_{n}\frac{\Delta_{n}(\omega)}{\sqrt{\omega^2-\Delta_{n}^{2}(\omega)}}
+i\Gamma_{nm}\frac{\Delta_{m}(\omega)}{\sqrt{\omega^2-\Delta_{m}^{2}(\omega)}}\right) 
\frac{\sqrt{\omega^2-\Delta_{n}^{2}(\omega)}}{\Delta_{n}(\omega)} \\
\label{SW2}
\phantom{\gamma_{n}}=Z_{n}(\omega)\sqrt{\omega^2-\Delta_{n}^{2}(\omega)}\>,
\end{eqnarray}
where in the last step we used (\ref{Eli1}) and $\Gamma_{nm}=1/(2\tau_{nm})$, $n,m=\sigma,\pi$, 
to replace the expression in the bracket in (\ref{SW1})
with $\phi_{n}(\omega)=\Delta_{n}(\omega)Z_{n}(\omega)$. Thus, $\gamma_{n}$ of \cite{Sung} is precisely 
equal to the quasiparticle energy $E_{n}(\omega)$, (\ref{ENERGY}), used in this work.  
The complex conductivity is obtained from the total response kernel $K_{n}(0,\nu)$ given by 
equations (38a) and (38b) in \cite{Sung} as $iK_{n}(0,\nu)/\nu$. One should keep in mind 
that as the cut along the real axis is crossed (i.e. $\omega$ is assumed to have 
infinitesimal negative imaginary part instead of infinitesimal positive imaginary part) 
$N_{n}(\omega)\rightarrow-N_{n}^{*}(\omega)$, $M_{n}(\omega)\rightarrow-M_{n}^{*}(\omega)$
and $E_{n}(\omega)\rightarrow-E_{n}^{*}(\omega)$ (see section VI of reference [9] in 
\cite{Sung} after which the derivation in \cite{Sung} was patterned). 
\noappendix
\newpage
\begin{figure}[b]
\includegraphics[angle=0,width=10cm]{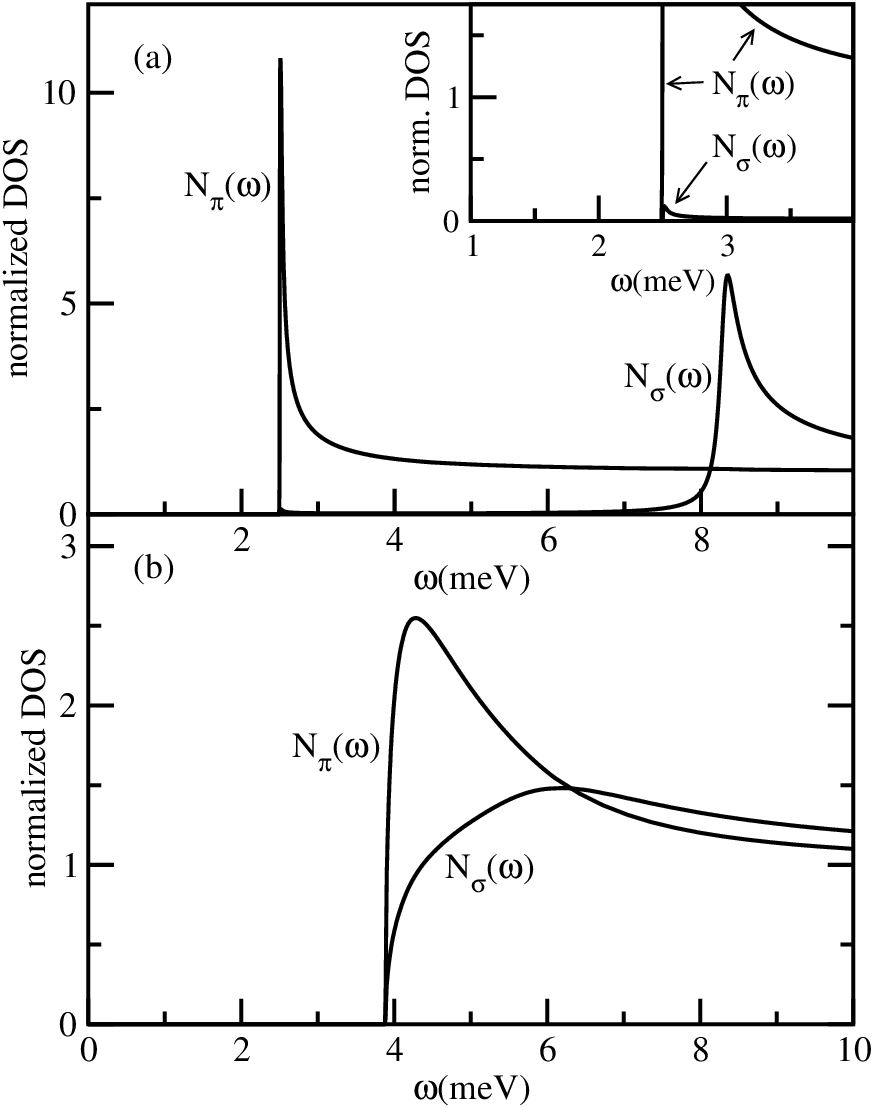}
\caption{The normalized quasiparticle densities of states $N_{\sigma(\pi)}(\omega)=\re[\omega/
\sqrt{\omega^2-\Delta_{\sigma(\pi)}^{2}(\omega)}$ for MgB$_{2}$ at T=5 K calculated using the
interaction parameters from \cite{Mitrovic1}. (a) The results obtained for the interband impurity scattering
rate 1/$\tau_{\pi\sigma}$=0.1$T_{c0}$, where $T_{c0}$ is the transition temperature of the clean
compound. The inset shows the details in the densities of states near the common gap. (b) The
results obtained for the interband impurity scattering rate 1/$\tau_{\pi\sigma}$=5$T_{c0}$ which
is used in the subsequent calculations of the frequency dependent optical conductivity.}
\label{fig:fig1}
\end{figure}
\newpage
\begin{figure}[b]
\includegraphics[angle=0,width=10cm]{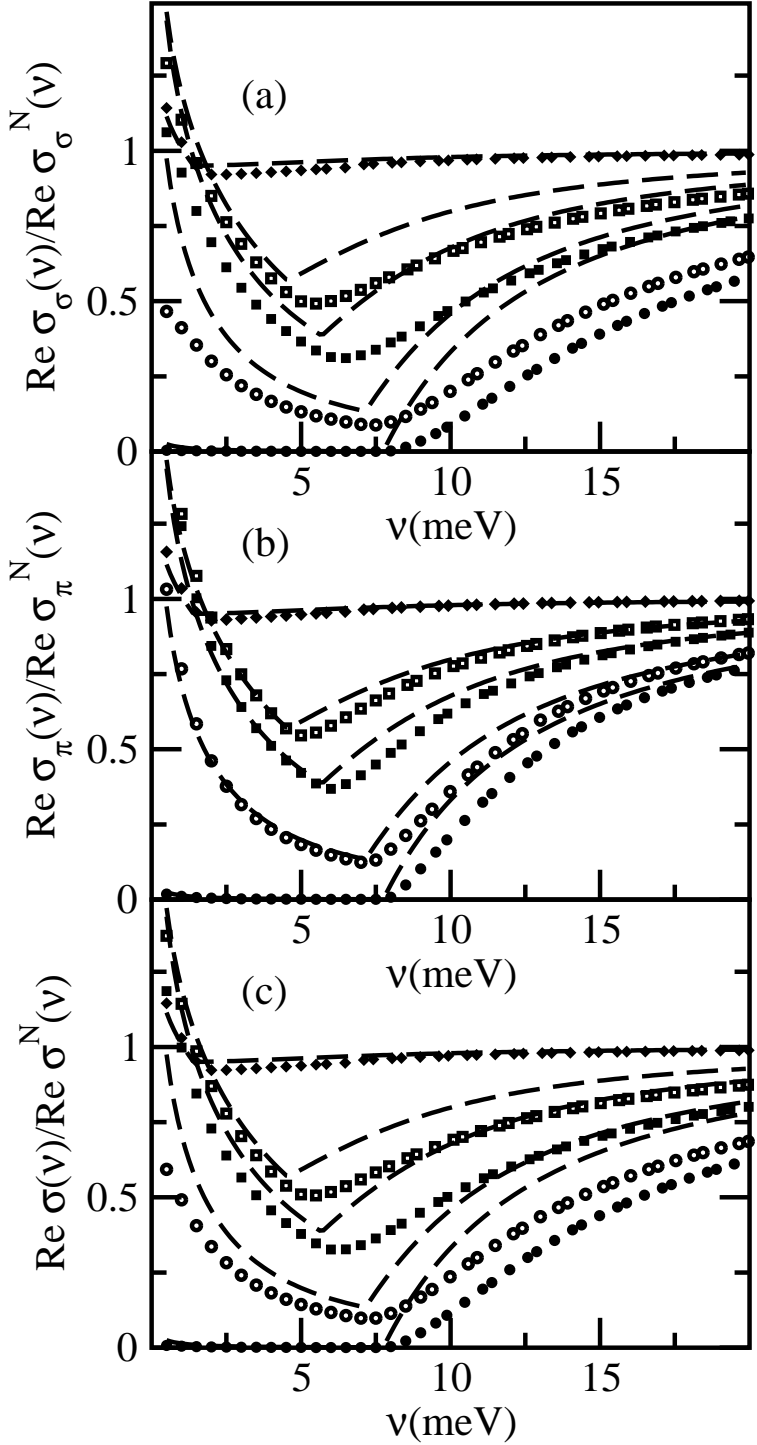}
\caption{Real part of the conductivity normalized to its normal state value at 40 for 
$T$= 7 K (filled circles), 19 K (opened circles), 26 K (filled squares), 29 K (opened 
squares) and 32.75 K (filled diamonds). The dashed lines are results obtained with 
Mattis-Bardeen theory at the same temperatures using $\Delta$=3.875 meV and assuming 
the BCS temperature dependence of the gap (the temperature increases from the bottom 
dashed curve to the top one). (a) Results for the $\sigma$-band conductivity. 
(b) Results for the $\pi$-band conductivity. (c) Results for the total conductivity 
$\sigma(\nu)=\sigma_{\sigma}(\nu)+\sigma_{\pi}(\nu)$.} 
\label{fig:fig2}
\end{figure}
\newpage
\begin{figure}[b]
\includegraphics[angle=0,width=10cm]{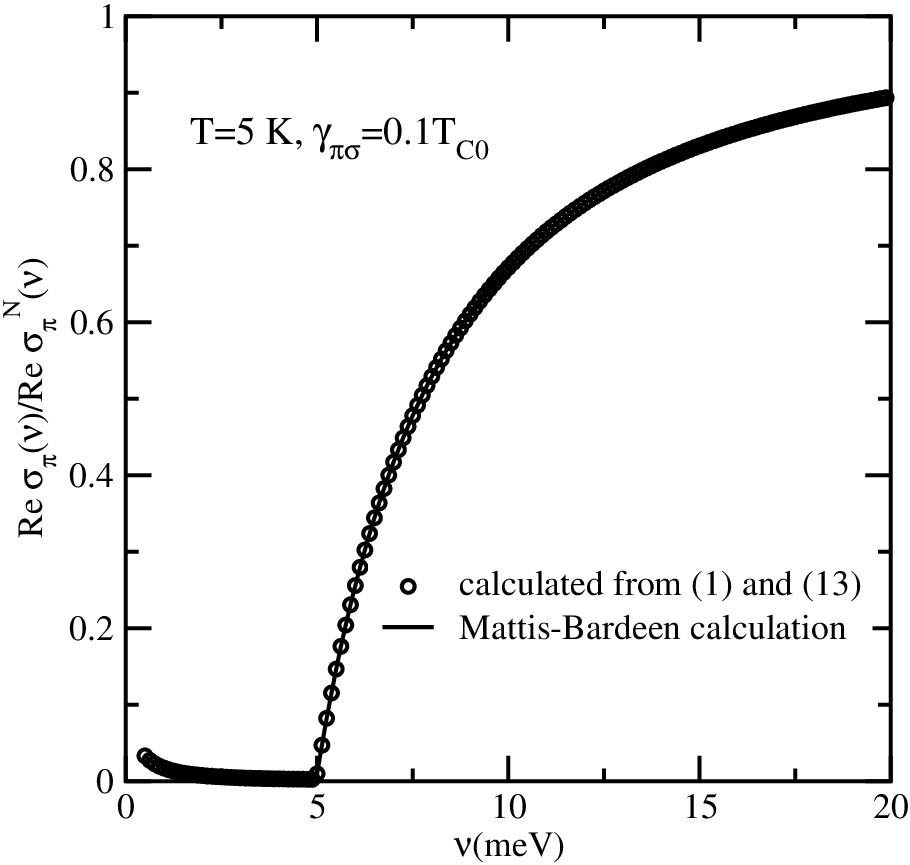}
\caption{Real part of the $\pi$-band conductivity at $T$=5 K normalized to its normal state value at 44 K.
The open circles are the results obtained from equations (\ref{conductivity}) and (\ref{nconductivity}) 
with the interband scattering rate $\gamma_{\pi\sigma}$=0.1$T_{c0}$ and with 
$\gamma_{\pi}=\gamma_{\pi\pi}+\gamma_{\pi\sigma}$ held at 2.6 eV. The solid line gives the 
results obtained with Mattis-Bardeen theory.}
\label{fig:fig3}
\end{figure}
\newpage
\begin{figure}[b]
\includegraphics[angle=0,width=10cm]{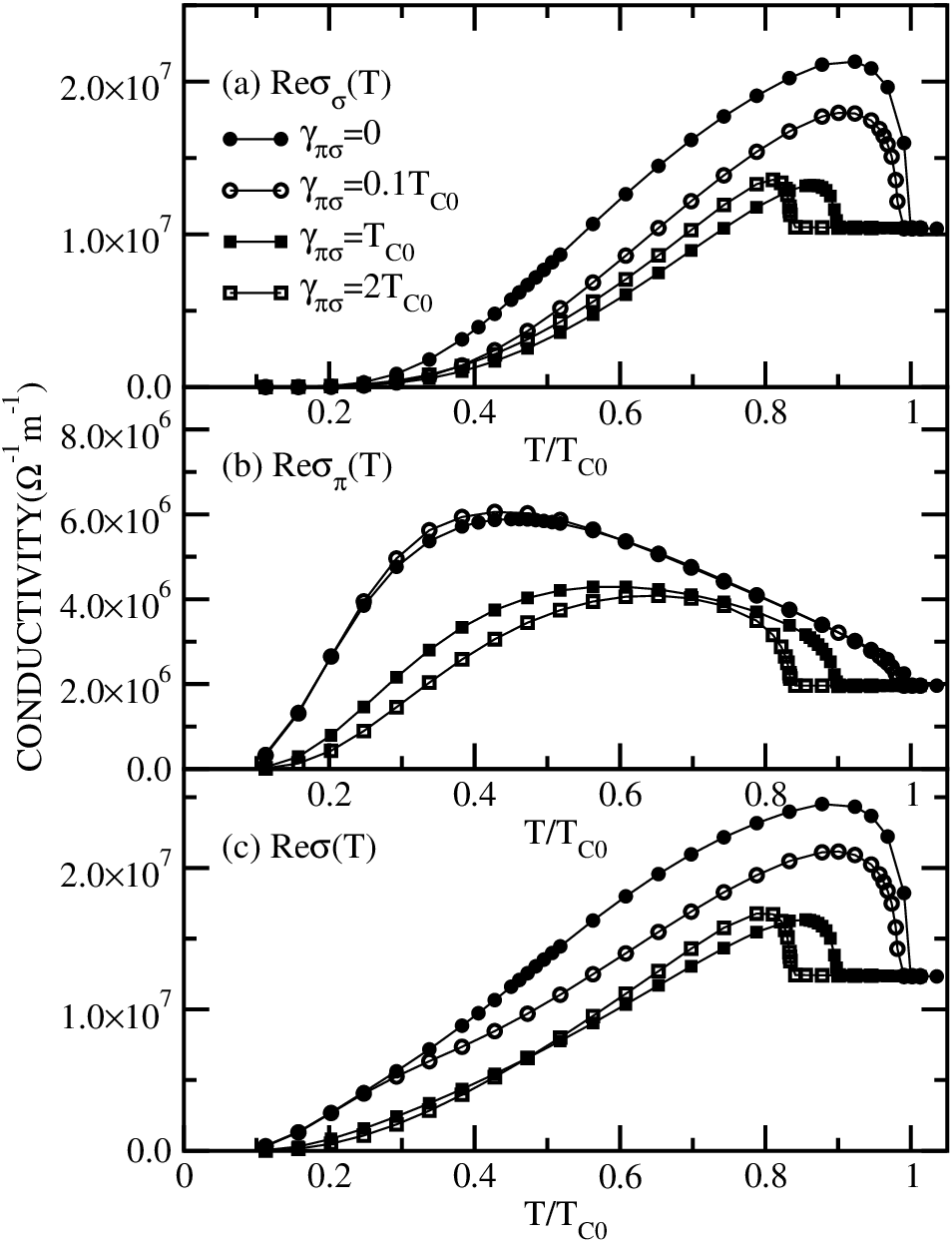}
\caption{The microwave conductivity at 8.5 GHz as a function of temperature for 
several different values of the interband scattering rate $\gamma_{\pi\sigma}$ in 
the units of transition temperature of the clean system $T_{c0}$, calculated using 
$\omega_{p\sigma}$=4.14 eV, $\omega_{p\pi}$=4.72 eV, $\gamma_{\pi}/\gamma_{\sigma}$=
7 [1] and $\gamma_{\pi}$=152.6 meV (see the text). (a) $\sigma$-band contribution 
$\re \sigma_{\sigma}(T)$ to the microwave conductivity. Different curves correspond 
to different values of $\gamma_{\pi\sigma}$ as indicated in the legend. (b) $\pi$-band 
contribution $\re \sigma_{\pi}(T)$ to the microwave conductivity. The identification 
of various curves is the same as in (a). (c) The real part of the total microwave 
conductivity $\re \sigma(T)=\re \sigma_{\sigma}(T)+
\re \sigma_{\sigma}(T)$. The meaning of different symbols is explained by the 
legend in (a).}
\label{fig:fig4}
\end{figure}
\newpage
\begin{figure}[b]
\includegraphics[angle=0,width=10cm]{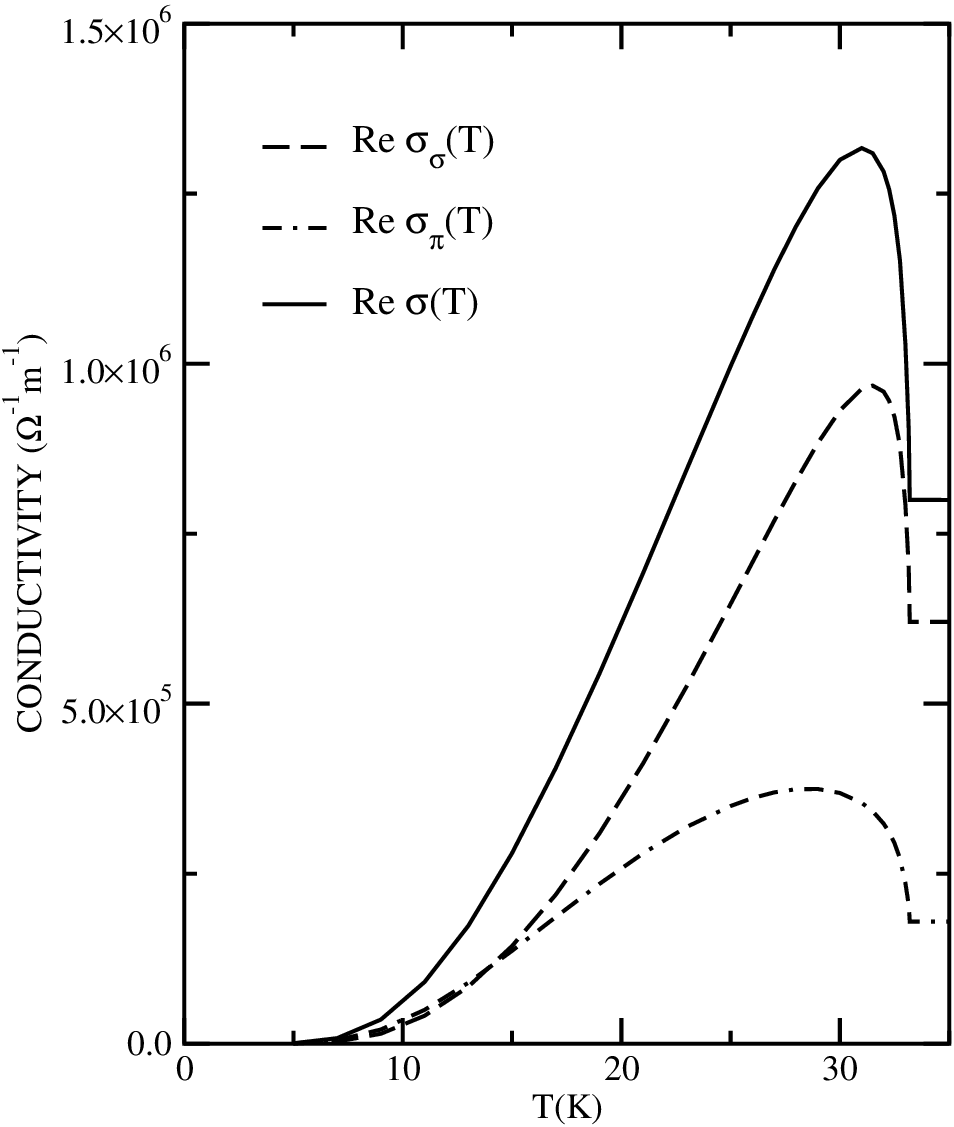}
\caption{Contributions of different bands and the total microwave conductivity 
at 8.5 GHz,
$\re \sigma(T)=\re \sigma_{\sigma}(T)+\re \sigma_{\sigma}(T)$, calculated for 
the same parameters used to obtain the infrared conductivities in figure 2: 
$\omega_{p\sigma}$=4.14 eV, $\omega_{p\pi}$=5.89 eV, $\gamma_{\pi}/\gamma_{\sigma}$=
7, $\gamma_{\pi}$=2.6 eV and $\gamma_{\pi\sigma}$=5$T_{c0}$. The corresponding 
quasiparticle densities of states at a low temperature are given in figure 1b.} 
\label{fig:fig5}
\end{figure}
\newpage
\begin{figure}[b]
\includegraphics[angle=0,width=10cm]{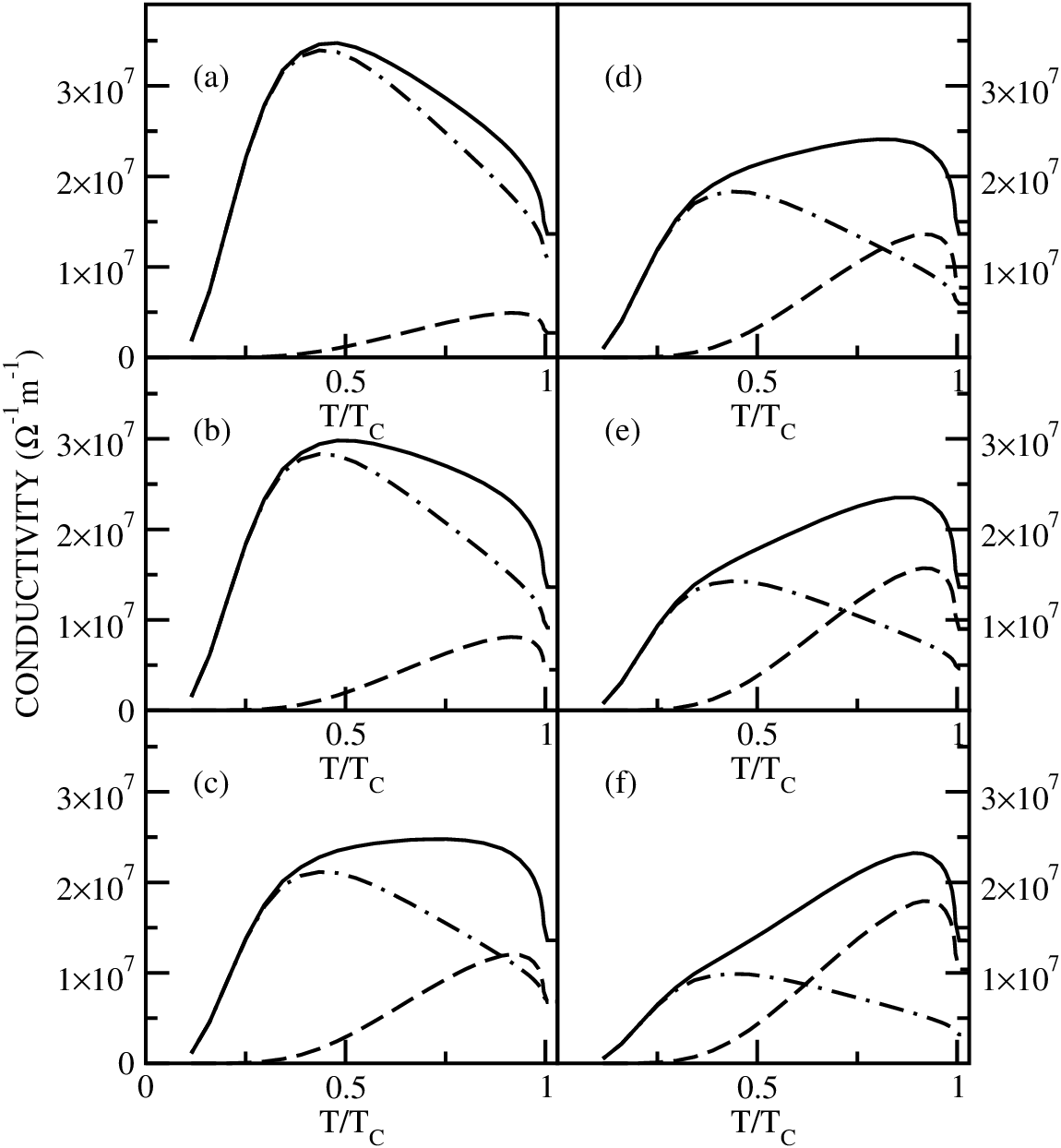}
\caption{Microwave conductivities for a fixed interband impurity scattering rate 
$\gamma_{\pi\sigma}$=0.1$T_{c0}$ for different values of the ratio $\gamma_{\pi}/\gamma_{\sigma}$ 
with $\gamma_{\pi}$ fitted to the measured microwave conductivity at $T_{c}$ of 1.37$\times$10$^{7}
/\Omega$m \cite{Jin} with $\omega_{p\sigma}$=4.14 eV, $\omega_{p\pi}$=5.89 eV.  
The contribution of the $\sigma$-band to the microwave conductivity is given by the dashed line, 
while that of the $\pi$-band is given by the dash-dotted line. The total microwave conductivity 
is represented by the solid line. 
(a) $\gamma_{\pi}/\gamma_{\sigma}$=0.5. (b) $\gamma_{\pi}/\gamma_{\sigma}$=1. 
(c) $\gamma_{\pi}/\gamma_{\sigma}$=2. (d) $\gamma_{\pi}/\gamma_{\sigma}$=2.67. 
(e) $\gamma_{\pi}/\gamma_{\sigma}$=4. (f) $\gamma_{\pi}/\gamma_{\sigma}$=6.67.}
\label{fig:fig6}
\end{figure}

\end{document}